\documentclass[twocolumn,amsmath,amssymb,pra]{revtex4}

\usepackage{graphics}
\usepackage{epsfig}
\usepackage{amsmath}
\usepackage{color}
\usepackage{dcolumn}
\usepackage{bm}
\usepackage{subfigure}

\newcommand{\A}{\mathcal{A}}
\newcommand{\Ket}[1]{\left|#1\right>}

\begin{document}

  \title{Ab-initio determination of Bose-Hubbard parameters for two 
         ultracold atoms\\
         in an optical lattice using a three-well potential}

       \author{Philipp-Immanuel Schneider, Sergey Grishkevich, and Alejandro Saenz}

       \affiliation{AG Moderne Optik, Institut f\"ur Physik,
         Humboldt-Universit\"at zu Berlin, Hausvogteiplatz 5-7,
         10117 Berlin, Germany}

       \date{\today}

  \begin{abstract}

   We calculate numerically the exact energy spectrum of the six dimensional 
   problem of two interacting Bosons in a three-well optical lattice.
   The particles
   interact via a full Born-Oppenheimer potential which can be adapted
   to model the behavior of the s-wave scattering length at Feshbach resonances.
   By adjusting the parameters of the corresponding Bose-Hubbard (BH) Hamiltonian
   the deviation between the numerical energy spectrum and the BH spectrum
   is minimized.
   This defines the optimal BH parameter set which we compare to the
   standard parameters of the BH model.
   The range of validity of the BH model with these parameter sets is
   examined, and an improved
   analytical prediction of the interaction parameter is introduced.
   Furthermore, an extended BH model and implications due to the energy
   dependence of the scattering length and couplings to higher Bloch
   bands at a Feshbach resonance are discussed.
  \end{abstract}

    \maketitle

\section{Introduction}
%
\label{sec:intro}

 The Hubbard model has its origin in the description of electrons in solids
 \cite{cold:hubb63}. However, 
 ultracold atoms in an optical lattice (OL) represent an almost perfect
 realization of this model with the additional advantage that many
 parameters such as the lattice depth and the interaction strength 
 can be controlled and characteristics of the system can be
 observed with high accuracy \cite{cold:bloc08}.
 The Bose-Hubbard model (BHM) is a special form of the Hubbard model
 describing Bosonic particles in an OL.
 Despite its striking simplicity 
 it is able to describe exciting phenomena such as the quantum phase transition 
 between the Mott insulator and superfluid phase which is determined by the 
 ratio $U/J$ of the interaction and the hopping parameter 
 \cite{cold:jaks98,cold:grei02a}. 
 If the system is quenched from one quantum phase to another, local many-body 
 relaxation effects appear even without thermal relaxation. They are
 explained by the coupling between neighboring
 lattice sites in the BH Hamiltonian \cite{cold:cram08}.
 Another prediction of the BHM is the existence of repulsively bound atom
 pairs in an optical lattice \cite{cold:vali08}. In contrast to a real solid,
 the optical lattice does not allow dissipation to phonons. As there is
 no resonant unbound state in the lattice, the repelling atoms are unable to
 decay. Evidence for this effect was found in an experiment with $^{87}$Rb
 atoms \cite{cold:wink06}.

 Not only large optical lattices but also few-well systems such as double and
 triple wells are frequently modeled by a BH Hamiltonian 
 \cite{cold:foel07,cold:chei08,cold:trot08}.
 These systems have potentially a large range of applications.
 The triple-well system was proposed to serve as a transistor, where the
 population of the middle well controls the tunneling of particles from the
 left to the right well~\cite{cold:stic07}.
 Double wells, effectively generated by two interfering optical lattices
 forming a superlattice, are promising candidates to implement one- and 
 two-qubit quantum gates \cite{cold:milb97,cold:ande06,cold:sebb06}.
 
 To link experimental data to theoretical predictions of the BHM
 and to finally realize practical applications such as quantum gates, 
 the precise knowledge of the parameters of the BH Hamiltonian is, however, important.
 We want to clarify in this work under which conditions the BHM is applicable
 and how the BH parameters may be improved.

 In order to examine and improve the validity of the BHM we calculate the energy
 spectrum of two interacting atoms in three wells of an OL
 by the help of a numerical routine introduced in Ref.~\cite{cold:gris09}.
 The reduction from the many-body problem to only two particles is
 senseful, because the BHM is often used to describe dilute quantum gases 
 with one or two atoms per lattice site.
 Also, the approximation of the OL by three wells does preserve all important
 microscopical features of the BHM of an OL. The BH Hamiltonian of the triple
 well allows for tunneling of particles between the sites with amplitude $J$, 
 for different onsite energies $\epsilon_i$ assigned to each of the
 sites $i=0,\pm 1$ and of course for an onsite interaction with strength $U$.
 The spectrum of the BH Hamiltonian of the triple well consists of six
 eigenenergies which depend on the parameter set
 $\mathcal P = \{J,U,\epsilon_0,\epsilon_{\pm 1}\}$. Although the parameters are
 predicted within the approximation of the BHM from the lattice depth,
 the confining potential, and the s-wave scattering length $a_{\rm sc}$, one
 can define an optimal set $\mathcal P_{\rm opt}$, which
 minimizes the deviation between the BH spectrum and the one obtained
 numerically. 

 In this work we identify mainly two sources which can lead to large discrepancies
 between the predicted parameters of the BHM and the optimal parameter set.
 First, effects of the additional potential which confines the particles
 in three wells are clearly visible in shallow lattices. Albeit this is an 
 artefact of the restriction to three wells it is, of course, important for the 
 theoretical description of double and triple wells by the BHM.
 The second and more severe discrepancy emerges if the scattering length
 $a_{\rm sc}$ is not significantly small compared to the characteristic trap
 length $a_{\rm ho}$. As the Wannier basis used in
 the BHM does not reflect interaction it leads to a wrong prediction of the
 interaction parameter $U$ for large scattering lengths as they 
 appear experimentally in the context of (magnetic) Feshbach resonances. 
 In this regime the known analytical solution for two particles in a 
 single harmonic well interacting via a point-like pseudopotential is often used to
 describe effects of the interaction \cite{cold:ties00,cold:stoc03,cold:koeh06,cold:ment09}.
 If no perturbation theory is implied, the harmonic approximation is, however, 
 limited to very deep lattices.
 In this work a simple-to-calculate correction factor to the harmonic
 approximation is introduced that extends the validity to shallower lattices
 and improves the prediction even for deep lattices.

 Not only for strong interaction the standard BHM faces a potential weakness. 
 For shallow lattices 
 particles in neighboring wells are able to interact and hopping of particles 
 to next-to-nearest neighbors can become important.
 One can, however, account for this by a so-called extended Bose-Hubbard model. 
 We examine in which regimes this extended model is necessary and useful.

 The paper is outlined as follows: In Sec.~\ref{sec:theory} we describe the
 Hamiltonian of the triple well and we give a
 short review of the theoretical description of ultracold atoms in optical
 lattices described by the BHM. In Sec.~\ref{sec:method} numerical
 methods and different approaches to determine and approximate the parameters
 of the BHM, especially the interaction parameter, are presented.
 Thereafter in Sec.~\ref{sec:results} the results of the full calculations and
 various approximations are compared.  Besides the values of the BH parameters
 and the resulting energy spectra,
 the influence of an energy-dependent scattering length is considered. 
 Furthermore, the behavior at the resonance of 
 the scattering length including couplings to higher Bloch bands is discussed.
 Finally, we study effects of next-neighbor interaction and hopping to 
 non-neighboring wells. 
 Conclusions are made in Sec.~\ref{sec:conc_outlook}.   

\section{System}
%
\label{sec:theory}

\subsection{Hamiltonian}

The two particles in the triple well are exemplarily represented
by two bosonic $^7$Li atoms in the electronic $a\,^3\Sigma_u^+$ state.
The Hamiltonian of this system in absolute coordinates is given by 
\begin{eqnarray}
   \label{eq:Ham0}
     H_{\rm full}(\vec{r}_1,\vec{r}_2)
           &=& - \frac{\hbar^2}{2m}\left( {\vec{\nabla}_1}^2 + {\vec{\nabla}_2}^2 \right) +
               V_{\rm int}(|\vec{r}_1-\vec{r}_1|)
           \nonumber \\
           &+& V_{\rm trap}(\vec{r}_1) +  V_{\rm trap}(\vec{r}_2)
\end{eqnarray}
where $V_{\rm trap}$ is the trapping potential and
$V_{\rm int}$ describes the $^7$Li-$^7$Li-interaction by the 
adiabatic Born-Oppenheimer (BO) potential of the $a\,^3\Sigma_u^+$ state.
This electronic state relates to a sample of spin-polarized atoms. 
The corresponding BO potential has the advantage
that it supports only few bound states which reduces the number of basis
functions needed to describe the wavefunction in the numerical calculation.

For an ideal (isotropic) OL the trapping potential is given as 
\begin{equation}
  V_{\rm OL}(\vec r\,) = \sum_{\xi=x,y,z} V_0 \sin^2(k_0 \xi)
\end{equation} 
where $V_0$ is the depth of the lattice and $a=\pi/k_0$ is its periodicity.
The wavelength $\lambda = 2 \pi / k_0$ is fixed to $ 1\,{\rm \mu m}$ in this work, 
which is in accordance to experiments with Li atoms in 
optical lattices \cite{cold:gros08}.
The harmonic approximation of the OL potential $V(\vec
r\,)=\frac{1}{2} m\omega^2r^2$ with $\omega^2 = 2 V_0 k_0^2/m$ 
defines the characteristic energy $\hbar\omega$
and the characteristic trap length $a_{\rm ho}=\sqrt{2\hbar/m\omega}$ 
which is a measure of the extent of the ground state solution 
in the harmonic approximation of the OL.
Another characteristic unit of the OL is the recoil energy 
$E_r =\hbar^2 k_0^2/2m$.

\begin{figure}[ht]
 \centering
      \includegraphics[width=0.45\textwidth]{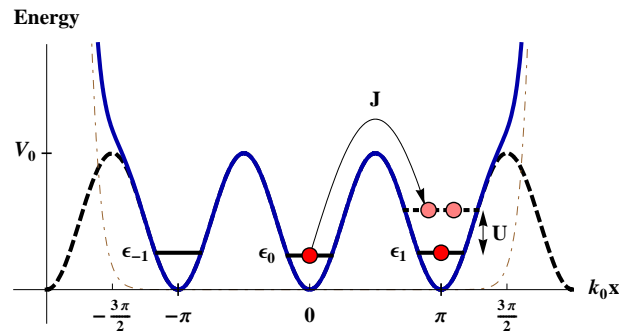}
 \caption{{\footnotesize (Color online)
 Visualization of the ``hopping''-parameter $J$, 
 the onsite energies $\epsilon_i\; (i=0,\pm 1)$
 and the interaction strength $U$. The potentials $V_{22}(\vec r\,)$ (blue solid),
 $V_{\rm OL}(\vec r\,)$ (black dashed) and $V_{\rm conf}(\vec r\,)$ (brown
 dash-dotted) are depicted for $y=z=0$.
}}
\label{fig:paramvis}
\end{figure}

To model a triple well, the OL potential is expanded to the twenty-second
order in $x$ direction and second order (i.e. harmonic approximation) in
$y$ and $z$ directions. This leads to a new potential $V_{22}(\vec r\,)$
consisting of three lattice sites which have in $x$-direction almost exactly
the form of the OL potential (see Fig.~\ref{fig:paramvis}).
The difference between the OL potential $\hat V_{\rm OL}$ and $\hat V_{22}$ can be 
regarded as an extra potential $\hat V_{\rm conf}= \hat V_{22}- \hat V_{\rm OL}$ 
which confines the particles in three sites of an infinite OL.
The triple-well Hamiltonian $\hat H_{\rm trip}$ for the two atoms $N=1,2$ is
thus given as 
\begin{equation}
\label{eq:HTrip}
  \hat H_{\rm trip} 
     = \sum_{N=1,2} \left(\frac{\hat p_i^2}{2m} + \hat V_{\rm OL}^{(i)}
                      + \hat V_{\rm conf}^{(i)}\right) 
       + \hat V_{\rm int}.
\end{equation}

In a real experimental setup the confinement to three wells could be either
due to a superlattice or could be generated by the
harmonic potential of an optical dipole trap which is, however, usually 
much shallower
than in our case where  $V_{\rm conf}(\vec r\,)$ rises rapidly for 
$|k_0 x| >\frac{3\pi}{2}$.

\subsection{Bose-Hubbard model}

\begin{figure}[ht]
 \centering
   \includegraphics[width=0.45\textwidth]{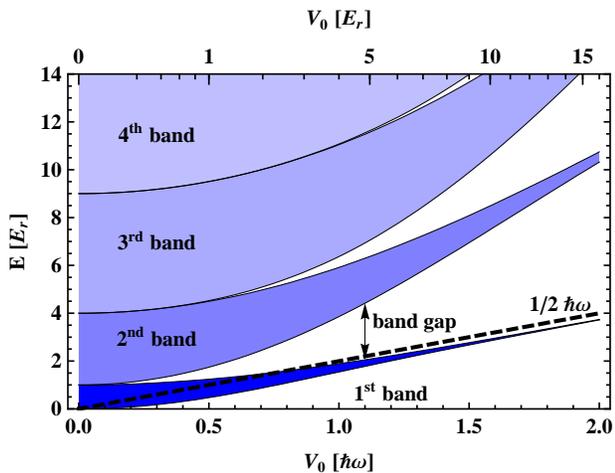}
 \caption{{\footnotesize (Color online) 
     Extents of the lowest four energy bands of the one-dimensional
     optical lattice as a function of the lattice depth $V_0$. 
     The energy of the first band
     can be approximated by the harmonic ground-state energy $\frac{1}{2}
     \hbar \omega$ (dashed line). A better approximation includes 
     a constant energy offset of $E_r/4$ (see the text).
}}
\label{fig:band-structure}
\end{figure}

The periodicity of the OL potential gives rise to band gaps in the dispersion
relation $E(\vec k\,)$. In contrast to a real solid, both the dispersion relation and the
Bloch solutions of the Hamiltonian $\hat H_{\rm OL} = \frac{\hat p^2}{2m}+\hat V_{\rm
  OL}$ are known analytically. 
The extents of the first four Bloch bands of a one-dimensional OL are depicted 
as a function of the lattice depth $V_0$ in Fig.~\ref{fig:band-structure}. 
For increasing lattice depths the band widths
shrink and the gaps between the bands increase. Especially, the gap between the
first and the second band prevents ultracold Bosons from occupying others than
the lowest Bloch band. The energy of states in this band can be roughly
approximated by the ground state energy $\hbar\omega/2$ of the harmonic approximation.
However, even for $V_0 \gg \hbar \omega$ there is a constant energy offset of $E_r/4$
between the harmonic approximation and the exact ground state energy.
This offset is a result of the anharmonicity of the lattice sites. It can be explained 
by a perturbation of the harmonic ground-state energy by the next-to-leading order expansion 
($\sim x^4$) of the OL potential~\cite{cold:slat52}.

The BHM makes use of the basis of Wannier functions, constructed with the aid
of the known Bloch 
solutions of the lowest band to approximate the triple-well Hamiltonian
$\hat H_{\rm trip}$. 
In the case of an infinite OL with no interaction this orthogonal basis spans 
completely the Hilbert space of the first Bloch band. In contrast to the Bloch
solutions, each Wannier function $w_j(\vec r\,)$ is localized at one lattice 
site $j$.
Wannier functions can be constructed in many different ways depending on the 
choice of phases of the Bloch solutions. In accordance to the widely used 
convention of the BHM we consider the form specified by Kohn in~\cite{cold:kohn59} 
which leads to maximally localized, real Wannier functions.

To describe systems with a finite number of wells, the amount of Bloch 
solutions is usually restricted to those which fulfill periodic boundary
conditions. For the triple-well 
potential used in this work, the boundaries may be set to 
$k_0 x_B = \frac{3 \pi}{2}$ or, in order to probe the outer 
potential walls, to $k_0 x_B = 2\pi$.
These conditions change slightly the shape of the Wannier
functions. 

We adapt the basis to the triple-well potential by using in $y$- and $z$-direction
the appropriate harmonic ground-state solution $h_0$ instead of 
the Wannier function for a one-well lattice. The adapted Wannier functions read
\begin{equation} 
 w_j(\vec r\,) =  w_j^{\rm 1D}(x)h_0(y)h_0(z).
\end{equation}
A detailed description of Wannier functions can be found in Refs.~\cite{cold:kohn59,cold:jaks05}.

Let $\hat b_j^\dagger$ ($\hat b_j$) be the bosonic operator of creation 
(annihilation) of a particle with the Wannier function
$w_j(\vec r\,)=\left<\vec r\,|w_j\right>$, then $\hat H_{\rm trip}$
(Eq.~\ref{eq:HTrip}) is written in the Wannier basis as
\begin{equation}
\begin{split}
  \hat H_{\rm Wan} =& \sum_{i,j}\left<w_i \right| \frac{\hat{p}^2}{2m} + \hat V_{\rm OL} +
  \hat V_{\rm conf}\left| w_j\right> \hat b^\dagger_i \hat b_j +\\
  \frac{1}{2}& \sum_{k,l,m,n}\left<w_k\right|\left<w_l\right|\hat V_{\rm
  int}\left|w_m\right>\left|w_n\right>  \hat b^\dagger_k \hat b_l^\dagger \hat
  b_m \hat b_n,
\end{split}
\label{eq:HamWan}
\end{equation}
with all indices running over $-1,0,+1$.

This is already the first approximation of the BHM as the Wannier basis 
is only complete for the first Bloch band if $\hat V_{\rm  int} = 0$ and 
$\hat V_{\rm conf} = 0$. The BHM further simplifies $\hat H_{\rm Wan}$  
by making the following assumptions: 
(i) The overlap between Wannier functions on different 
lattice sites is small. This reduces the $i,j$-summation to diagonal 
elements $i=j$ and neighboring lattice sites $\left<i,j\right>$.
(ii) The confinement potential $\hat V_{\rm conf}$ varies slowly and does not couple
Wannier functions at different lattice sites, i.~e. 
$\left<w_i \right| \hat V_{\rm conf}\left| w_j\right>=0$ for $i \neq j$.
(iii) The interaction is short-ranged and only takes place between particles
in the same well.

Applying these assumptions, the final BH Hamiltonian takes the form
\begin{equation}
  \hat H_{\rm BH} = - J\sum_{<i,j>} \hat b^\dagger_i \hat b_j + 
   \frac{U}{2} \sum_{i}\hat b^\dagger_i \hat b_i^\dagger \hat b_i \hat b_i + 
   \sum_i \epsilon_i \hat b^\dagger_i \hat b_i,
\label{eq:BHHam}
\end{equation}
with the hopping parameter 
$J = - \left<w_0 \right| \frac{\hat{p}^2}{2m}+ \hat V_{\rm OL} \left| w_{1}\right>$,
the onsite energies 
$\epsilon_{i} = \left<w_i\right| \frac{\hat{p}^2}{2m}  + \hat V_{\rm OL} + 
\hat V_{\rm conf}\left| w_i\right>\, (i = 0,\pm 1)$ 
and the interaction parameter
$U = \left<w_0\right|\left<w_0\right|\hat V_{\rm
  int}\left|w_0\right>\left|w_0\right>$. 
Even though the BH Hamiltonian has a simple form, there does not exist a
general solution for large lattices with many particles.
However, for two particles in a triple well the BH Hamiltonian is easily 
diagonalizable.

The integrations occuring in the definition of the BH parameters are performed 
consistently within the periodic boundary
conditions of $x$. The Bloch functions which are used to construct the
Wannier functions are also normalized to one within the boundaries. 
Since the trapping potential is symmetric, $\epsilon_{-1}=\epsilon_{+1}$ holds.
For a visualization of the BH parameters see Fig.~\ref{fig:paramvis}.

A further important approximation, which is very common in the physics of
ultracold atomic gases, is the replacement of the full BO interaction potential 
by a point-like pseudopotential~\cite{cold:busc98}
\begin{equation}
\label{eq:V_pp}
  V_{\rm ps}(\vec r\,)=\frac{4 \pi \hbar^2 a_{\rm sc}}{m}\delta^{(3)}(\vec r\,) \frac{\partial}{\partial r} r
\end{equation}
that depends only on the s-wave scattering length $a_{\rm sc}$.
This simplifies the integration for the interaction parameter drastically to
\begin{equation}
\label{eq:U_BH}
  U = \frac{4 \pi \hbar^2 a_{\rm sc}}{m} \int {\rm d}^3 \vec r\,\left| w_0(\vec r\,)\right|^4.
\end{equation}

The formulations of $J,U,\epsilon_0,$ and $\epsilon_{\pm 1}$ described above
are in the following denoted as the ``BH representation'' of the
parameters and will be indicated with the superscript ``BH''. 

It should be noted that not only the Wannier basis leads to the form of the BH Hamiltonian 
(Eq.~\ref{eq:BHHam}). Any other nearly orthogonal basis of functions
localized in one of the wells leads to the same form, with only different
values of the parameters.
The usual basis of the BH Hamiltonian of two particles in a triple well 
is the occupation-number basis $\{\Ket{002},\Ket{011},\Ket{101},\Ket{020},
\Ket{110},\Ket{200}\}$ which specifies 
the amount of particles in the left, middle and right well.
In this respect it is even possible to think of a two-particle instead of a one-particle
basis to arrive at the BHM. As a result, the BH Hamiltonian is principally
able to describe strongly interacting particles without the need to expand the 
full Hamiltonian (\ref{eq:Ham0}) in a large basis of one-particle Wannier
functions of several Bloch bands.
It is hence natural to imagine that there is a basis for the BHM which is far
better adapted to an OL with interacting particles in the presence of a 
confining potential $\hat V_{\rm conf}$ than the Wannier basis. Although this
optimal basis is unknown, it allows us to vary
the BH parameters in order to adapt the model to the full numerical solution 
as is described below.

\section{Method}
%
\label{sec:method}
\subsection{Numerical calculation in the triple well}

To perform the full numerical calculation of the eigenenergies and eigenfunctions of
Hamiltonian $\hat H_{\rm trip}$ the approach described in \cite{cold:gris09} is
adopted. Briefly, in this method the lattice potential is expanded into a
Taylor series and the resulting Hamiltonian is written as a  
sum of center-of-mass (COM) Hamiltonian $\hat H_{\rm COM}$, relative-motion
(REL) Hamiltonian $\hat H_{\rm REL}$ and coupling terms $\hat W$.
The eigenfunctions and eigenvalues of $\hat H_{\rm COM}$ and $\hat H_{\rm REL}$
are calculated separately in a basis of spherical harmonics
$Y_l^m(\theta,\phi)$ times  
B splines $B_{\alpha}(r)$. The position of knots of the B splines in REL can be
adjusted to adequately describe both the long range behavior in the trap 
($\sim 10^4 a_0$) and the short range behavior due to the molecular
(interatomic) interaction ($\sim 65 a_0$).
The full solution including the coupling $\hat W$ is calculated from
the eigenfunctions of $\hat H_{\rm COM}$ and $\hat H_{\rm REL}$ by the 
configuration-interaction (CI) method which is also known as exact diagonalization. 
The code is adapted to the symmetry group $D_{2h}$ of a cubic
three-dimensional OL, which simplifies the calculation and 
allows to reduce the REL basis to functions which are symmetric under inversion
and describe therefore Bosonic particles.

In Ref.~\cite{cold:gris09}, a Taylor expansion up to the sixth order of the OL
is used to examine the two-particle interaction in a single well.
The ground state was described by a basis which consisted of angular components
up to $l=3$.
The triple well considered in the present work demands much higher angular
components owing to its high degree of anisotropy. 
We need an expansion up to $l=32$ and about 70 B splines in COM and an expansion
up to $l=30$ and 130 B splines in REL to obtain
converged solutions for lattice depths up to $V_0=12 E_r$.
Altogether, about $16\ 000$ basis functions are used to describe the REL
eigensolutions and $11\ 000$ to describe the COM eigensolutions. 
From these solutions about $10\ 000$ configurations form the basis of the CI 
calculations.

To tune the  s-wave scattering length $a_{\rm sc}$ to values in
the range $a_{\rm sc} \in (-\infty,+\infty)$ the inner wall of the BO
potential is slightly shifted as suggested in Ref.~\cite{cold:gris09}.
This procedure can move the least bound state close to the dissociation
threshold or leads to the creation of a new bound state.
Simultaneously the scattering length grows to $+\infty$ or $-\infty$, respectively, 
which offers a knob to vary $a_{\rm sc}$ in a controlled way.

\subsection{Numerical Determination of the optimal BH parameters}
The numerical determination of the optimal BH parameter set is done by 
fitting the parameters $\mathcal{P} = (J,\epsilon_0,\epsilon_{\pm 1},U)$ 
such that an optimal agreement between the eigenenergies of the BH Hamiltonian 
and the full numerical solution is achieved.

In more detail, let 
\begin{equation}
  \hat H_{\rm trip} \Ket{\Psi_i} = E_i \Ket{\Psi_i}
\end{equation}
 be the stationary solution of the triple-well Hamiltonian
and
\begin{equation}
  \hat H_{\rm BH}^{\mathcal P} \Ket{\Phi_j^{\mathcal P}} = 
  \mathcal{E}_j^{\mathcal P} \Ket{\Phi_j^{\mathcal P}} 
\end{equation}
be the stationary solution of the BH Hamiltonian $\hat H_{\rm BH}^{\mathcal P} \equiv \hat
H_{\rm BH}$ with parameter set $\mathcal{P}$. Due to the restriction to one
basis function per particle and lattice site, the second eigenvalue equation has
only six solutions with energies $\mathcal{E}_1^{\mathcal P},\dots,\mathcal{E}_6^{\mathcal P}$.

With the aid of the eigenenergies, we obtain the optimal set of BH parameters 
$\mathcal P^{\rm opt}$ by minimizing the measure
\begin{equation}
  f(\mathcal P) = \sum_{i=1}^6\left(\frac{E_i - \mathcal{E}_i^{\mathcal P}}{E_i} \right)^2.
\end{equation}

However, a fit of all four parameters at once is problematic. 
Since only six eigenenergies exist, a local minimum of $f(\mathcal P)$ may be
found that leads to a non-optimal parameter set $\mathcal P$.
On the other hand, it is inherent to the BHM, that all effects
due to interaction are contained in the interaction parameter $U$ leaving the
other parameters unaffected.
Therefore, we determine the optimal parameter set in a two-step process:
(i) The parameters $J,\epsilon_0,\epsilon_{\pm 1}$ are fitted
    to match the energies of a numerical calculation without interaction.
(ii) The parameter $U$ is fitted to match the
energies of a numerical calculation \emph{with} interaction, keeping the other 
BH parameters fixed to the values of the previous fit. 

\subsection{Estimation of the interaction parameter $U$}
\label{ssec:Ucorr}
 As already mentioned, the Wannier functions do not reflect effects of interaction.
 Being a one-particle basis, the Wannier basis is only complete for
 non-interacting particles. Thus, one cannot assume that in strong interacting 
 regimes the BH representation of the interaction parameter given in
 Eq.~(\ref{eq:U_BH}) is applicable and one should estimate the energy offset
 caused by interaction differently.

\begin{figure}[ht]
 \centering
   \includegraphics[width=0.44\textwidth]{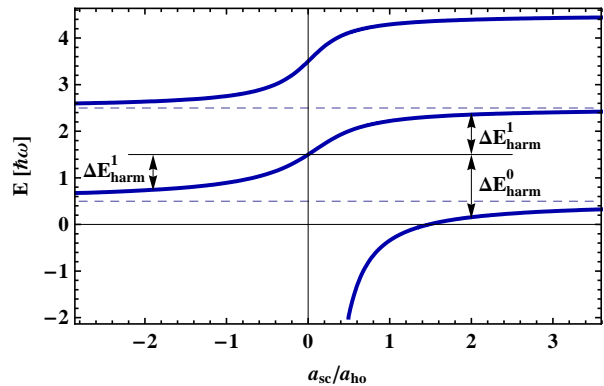}
 \caption{{\footnotesize (Color online)
 Analytical energy spectrum of relative-motion $l=0$ states of two particles 
 interacting by the potential $\hat V_{\rm ps}$ (Eq.~\ref{eq:V_pp}) in a harmonic trap
 as a function of the scattering length. 
 The spectrum defines the energy offsets $\Delta E_{\rm harm}^0$ and 
 $\Delta E_{\rm harm}^1$ relative to the ground-state 
 energy in the absence of interaction, $\frac{3}{2}\hbar\omega$.
 }}
\label{fig:harmenergies}
\end{figure}

 Fortunately, an analytical solution is known for a system of two identical 
 particles in an isotropic or anisotropic
 harmonic trap interacting only via the pseudopotential of Eq.~(\ref{eq:V_pp})
  \cite{cold:busc98,cold:idzi05}.
 While for no interaction ($a_{\rm sc}=0$) the eigenenergies in the isotropic
 case are given as
 $E_{n, l} = \hbar\omega(\frac{3}{2} + 2n + l)$, a non-vanishing scattering
 length leads to a shift of all eigenenergies with $l=0$. The dependence of the
 energies $E_{n, l=0}$ on the scattering length is shown in
 Fig.~\ref{fig:harmenergies} and is given by the solution of the implicit equation
\begin{equation}
  \frac{a_{\rm sc}}{a_{\rm ho}} = 
  \frac{1}{2} \tan\left(\frac{\pi \epsilon}{2} + \frac{\pi}{4}\right) 
  \frac{\Gamma\left( \frac{\epsilon}{2} + \frac{1}{4}\right)}{
  \Gamma\left( \frac{\epsilon}{2} + \frac{3}{4}\right)},
\label{eq:HarmEner}
\end{equation}
 where $\epsilon = E/\hbar\omega$ and $\Gamma$ denotes the Gamma function
 \cite{cold:bold02}. 
 Additionally, for positive scattering lengths one bound
 eigensolution appears (lowest branch).
 For ultracold atoms only the first two states with energies lower than $E_{0,
   1}=\frac{5}{2}\hbar\omega$ should be occupied. This corresponds to the
 reduction to the first Bloch band in the OL.

 The known energy spectrum offers the possibility to estimate the energy offset
 due to interaction as a function of the scattering length. As shown in
 Fig.~\ref{fig:harmenergies}, for positive scattering lengths one has the 
 choice between two
 branches; one with a negative offset $\Delta E_{\rm harm}^0$ to the lower
 branch and one with a positive  offset $\Delta E_{\rm harm}^1$ to the upper
 branch. For negative $a_{\rm sc}$ only the latter branch exists and $\Delta E_{\rm
   harm}^1$ is negative.

Since the interaction of the BHM has purely onsite character, one can expect 
that these energy offsets provide a good first approximation  $U^{\rm
  harm}=\Delta E_{\rm harm}^{0,1}$
of the interaction parameter for strong interaction. Depending on which
offset is chosen lower or higher parts of the energy spectrum may be modeled.
The confinement in the harmonic trap is, however, stronger than in a sinusoidal OL.
Thus, one can expect that the harmonic approximation tends to overestimate the
strength of the onsite interaction in the OL.

Despite the weakness of the Wannier basis for strong interaction it can
offer crucial information on anharmonic effects.
Its completeness for $\hat V_{\rm int}= 0$ and $\hat V_{\rm  conf}= 0$ allows
an almost exact prediction of the behavior of $U$ for $a_{\rm
  sc}\rightarrow 0$. (To first order, $\hat V_{\rm conf}$ has no influence on
$U$.)
This observation can be used to define a correction factor $\A$ to the
harmonic approximation which accounts for anharmonic effects.
The factor is defined by demanding that $\A \cdot \Delta E_{\rm  harm}^1$
has the same linear behavior as $U^{\rm BH}$ for $a_{\rm  sc} \rightarrow 0$, 
i.\,e. 
\begin{eqnarray}\label{eq:Corr}
  \A \cdot \left( \frac{\rm d }{ d a_{\rm sc}} \Delta E_{\rm harm}^1
  \right)_{a_{\rm sc}\rightarrow 0} \stackrel{!}{=} \frac{{\rm d} U^{\rm BH}}{d a_{\rm
      sc} } \nonumber \\
  \Rightarrow \A = \frac{\sqrt{\pi} a_{\rm ho}}{2\hbar\omega} \cdot \frac{4 \pi \hbar^2}{m} 
                   \int {\rm d}^3 \vec r\,\left| w_0(\vec r\,)\right|^4.
\end{eqnarray}

\begin{figure}[ht]
 \centering
   \includegraphics[width=0.44\textwidth]{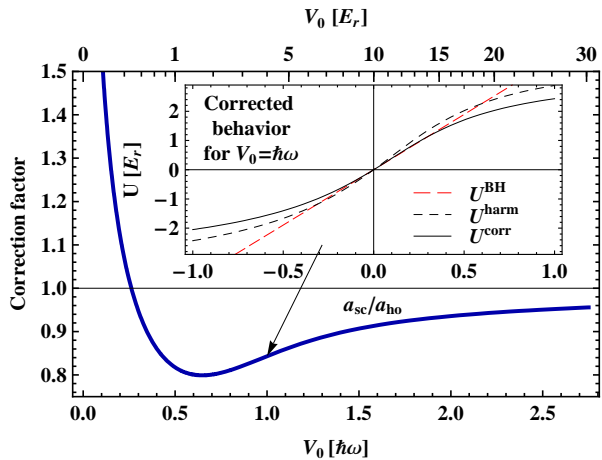}
 \caption{{\footnotesize (Color online) The correction factor $\A$ as a
     function of the lattice depth $V_0$.
{\bf Inset:} Comparison of $U^{\rm BH}$, $U^{\rm harm}$, and $U^{\rm corr}$ 
 for $V_0=\hbar\omega$ as a function of the scattering length. 
 In accordance with the definition of $\A$ the behavior of $U^{\rm BH}$ and
 $U^{\rm corr}$ matches for small $a_{\rm sc}$.  
 }}
\label{fig:Corr}
\end{figure}

Remarkably, the factor $\A$, which is shown in Fig.~\ref{fig:Corr} for
different lattice depths $V_0$, approaches even for very deep lattices only
slowly unity. Thus it is needed to correct the harmonic approximation even 
for very deep lattices. For shallow lattices the factor 
rises above one and goes for $V_0\rightarrow 0$ to $+\infty$. 
This is due to the fact that periodic boundary conditions 
are implied. Thus, the normalization of the
Wannier functions within the boundaries prevents the integral in the BH
representation of $U$ to vanish for $V_0\rightarrow 0$ while $a_{\rm
  ho}/\hbar\omega$ goes to infinity. 

The validity regime of the corrected harmonic approximation 
$U^{\rm corr}\equiv \A U^{\rm harm}$ depends on the applicability of the
pseudopotential approximation with an energy-independent scattering length
used in $U^{\rm harm}$ and the validity range of $\A$. The latter can be
roughly predicted by considering that the correction factor accounts for the
different confinement in the harmonic trap and the real OL for a certain
ground state energy $E$ of noninteracting particles.
In order for $\A$ to be valid for $a_{\rm sc}\neq 0$, the deviation of the energy 
($\approx U$) should be small compared to the depth of the lattice $V_0$, 
which simply means
\begin{equation}
\label{eq:validity_of_correction}
 |U| \approx |U^{\rm corr}| \ll V_0.
\end{equation}
This is to some extent counter-intuitive as for a fixed scattering length
$a_{\rm sc}$ an increase of $V_0$ is accompanied by an increase of $|a_{\rm sc}/a_{\rm ho}|$.
This leads to a breakdown of the BH representation and to errors caused by the
use of an energy-independent scattering length in $U^{\rm harm}$.
Since at the same time the error of the correction factor $\A$ shrinks it is
important to quantify and compare the different errors in order to understand 
when $U^{\rm corr}$ is applicable.

\subsection{Sextic approximation} 

An alternative way to look at effects of anharmonicity on the interaction is to consider 
an optical lattice not expanded to second or twenty-second but to the
sixth order in $x$-direction. This sextic trap consists of one well, but
includes anharmonic effects and coupling between relative and center-of-mass
motion~\cite{cold:gris09}. The energy difference of two atoms in a sextic well
with and without interaction defines the interaction parameter $U^{\rm sext}$.

\subsection{Energy-dependent scattering length}

The scattering length used in the pseudopotential (Eq.~\ref{eq:V_pp}) 
is defined by the s wave of a trap-free scattering
process in the limit of zero collision energy. The solution in the trap has,
however, a different asymptotic behavior and a finite ground-state energy.
Much work has been devoted to the question when a pseudopotential can be used 
in a trap and which scattering length should be chosen for energies greater 
than zero \cite{cold:bgao98,cold:ties00,cold:bloc02,cold:bold02}.  
It has been shown that the use of an energy-dependent effective scattering
length can largely extend the range of validity of the pseudopotential
approximation towards strong confinement and strong interaction in harmonic
traps \cite{cold:blum02,cold:bold02,cold:gris09}.

To quantify the impact of the energy dependence independently of anharmonic
effects we consider two $^7$Li atoms in a harmonic trap and determine 
numerically the ground-state energy. Plugging this energy into the right-hand side 
of Eq.~(\ref{eq:HarmEner}) defines the ``optimal'' scattering length $a_{\rm
  opt}$ for the description of the interaction by the pseudopotential.
In its range of validity the effective scattering length is equal to the
optimal scattering length.

The numerical calculation with the full BO interaction in an OL makes it possible to
compare the effect of an energy-dependent scattering length to effects caused
by the anharmonicity of the OL and the incompleteness of the Wannier basis.

%
\section{Results}
%
\label{sec:results}

This section is devoted to a comparison of 
the analytical approaches to describe interacting particles in a triple well 
with the full numerical results which determine the optimal BH parameters. 
First, the interaction-independent parameters $J,\epsilon_0,\epsilon_{\rm \pm 1}$ are
considered for a varying lattice depth $V_0$.
Then the discussion is extended to interacting particles for both different lattice
depths and different scattering lengths. 

Here a short overview of the abbreviations used in the following.

\vspace{0.25cm}
\begin{tabular}{l p{0.35\textwidth}}
{\bf BH} & Parameters in BH representation and eigenenergies of the BHM with
these parameters. \\
{\bf harm} & The same as ``BH'' but with $U^{\rm harm}$ determined 
           in the harmonic trap. \\
{\bf corr} & The same as ``BH'' but with the corrected $U^{\rm corr}=\A \cdot U^{\rm
            harm}$. \\
{\bf opt} & The optimal BH parameters and eigenenergies of the BHM with these
            parameters. This abbreviation is also used for the optimal scattering length.\\
{\bf sext} & Interaction parameter $U^{\rm sext}$ obtained from the sextic
             approximation of the OL.\\
{\bf num} & The eigenenergies of the full numerical CI calculation.
\end{tabular}
\vspace{0.25cm}

\subsection{Onsite energies} 
\label{ssec:onsite_energies}

\begin{figure}[ht]
 \centering
   \includegraphics[width=0.45\textwidth]{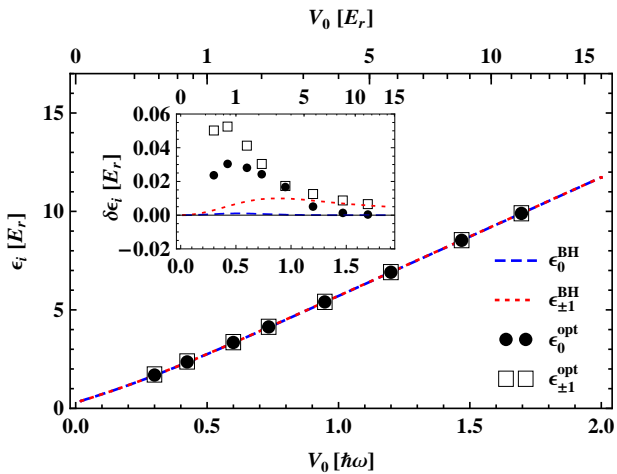}
 \caption{{\footnotesize (Color online) Onsite energies as a function of the lattice
  depth $V_0$: the optimal values together with the BH representation 
  for the middle well ($\epsilon_0$) and the outer wells ($\epsilon_{\pm 1}$) for
  $k_0 x_B = 3\pi/2$.\\
  {\bf Inset:} In order to examine the impact of the confining potential $\hat V_{\rm conf}$ 
  the value of $\delta\epsilon_i = \epsilon_i - \left<w_0\right|\frac{\hat{p}^2}{2m}  
  + \hat V_{\rm OL}\left|w_0\right>$ is depicted, where the value of the onsite energy 
  for no confinement is subtracted.
  Clearly, for $V_0>1.5\,\hbar \omega$ both results are converged.
}}
\label{fig:onsite}
\end{figure}

  The optimal onsite energies for the middle well and the outer wells are shown in
  Fig.~\ref{fig:onsite}. On the first glance, they seem to be very well matched by the 
  corresponding BH representations. However, by subtracting the value of the onsite 
  energy for no confinement 
  $\left<w_0\right|\frac{\hat{p}^2}{2m}  + \hat V_{\rm OL}\left|w_0\right>$ in the
  inset of  Fig.~\ref{fig:onsite} the impact of $\hat V_{\rm conf}$ on the BH 
  representation and the optimal values of the onsite energies can be examined.
  Generally, the shallower the lattices the more the optimal onsite energies 
  of both wells are lifted by the confinement. Only for $V_0>\hbar\omega$ this trend 
  is reflected by the BH representation, while for shallower lattices boundary conditions 
  at $k_0 x_B = 3\pi/2$ underestimate the impact of the confinement.   
  On the other hand, for $k_0 x_B = 2\pi$ (not shown in the graph) the boundary reaches deep
  inside the walls of the confinement and the BH representation
  extremely overestimates $\epsilon_{\pm 1}$ for $V_0<1.7\,\hbar\omega$ and even 
  $\epsilon_{0}$ for $V_0<0.8\,\hbar\omega$.

  A reason for the inaccuracy of the BH representation of
  $\epsilon_0$ and $\epsilon_{\pm 1}$ is that only for a slowly varying confining potential
  $V_{\rm conf}(\vec r\,)$ the Wannier functions form a good basis. 
  As $V_{\rm conf}(\vec r\,)$ varies strongly for $|k_0 x| > \frac{3 \pi}{2}$
  the lattice must be considerably deep so that the particles do not ``see'' the form 
  of the boundary and the distance $x_B$ of the boundary conditions must be sufficiently small 
  not to reach inside the region of the strong increase of the confining potential 
  (see Fig.~\ref{fig:paramvis}).  For boundary conditions at $k_0 x_B = 3\pi/2$ and
  lattice depths $V_0 > 1.5\,\hbar\omega$ both the values of the onsite energies and 
  their two orders of magnitude smaller differences are well described by the 
  BH representation.
  It can be assumed that for larger lattices with smooth harmonic trapping
  potentials, the BHM converges faster. 

\subsection{Hopping parameter} 
\label{ssec:hopping_parameter}

  As in the case of the onsite energies, the confining potential can cause also
  problems for the prediction of the hopping parameter.
  Fig.~\ref{fig:hopping} (a) shows the optimal hopping parameter and its BH
  representation for different boundary conditions $k_0 x_B = 3\pi/2,
  2\pi$ and $\infty$. One can see that the choice of
  the boundary conditions influences considerably the BH representation for
  shallow lattices. It seems that in this special case boundaries at $2\pi$ or
  infinity produce the best prediction. However, similarly to the case of the onsite
  energies all BH representations predict the correct hopping parameter for
  $V_0>1.5\,\hbar\omega$. 
  The reason is analogous. The BHM assumes that $\hat V_{\rm
    conf}$ is approximately constant within one well. 
  Therefore, due to the orthogonality of the  Wannier functions, 
  $\left<w_0 \right|\hat V_{\rm conf} \left| w_1 \right>=0$ holds.
  This is essential for the definition of the hopping parameter, where the
  influence of the confinement potential is neglected.
  However, as described in the last paragraph, for shallow lattices the form of
  the boundary becomes important and strongly varying parts of the potential
  are probed. One can expect an error of the BH representation for $J$ of
  the size $\Delta J = \left<w_0 \right|\hat V_{\rm conf} \left| w_1
  \right>$ which is negligible only for sufficiently deep lattices.

\subsubsection*{Energy spectrum for no interaction}  

\begin{figure}[ht]
 \centering
  {\bf (a)}\begin{minipage}[t]{0.45\textwidth}
     \vspace{0cm}\includegraphics[width=\textwidth]{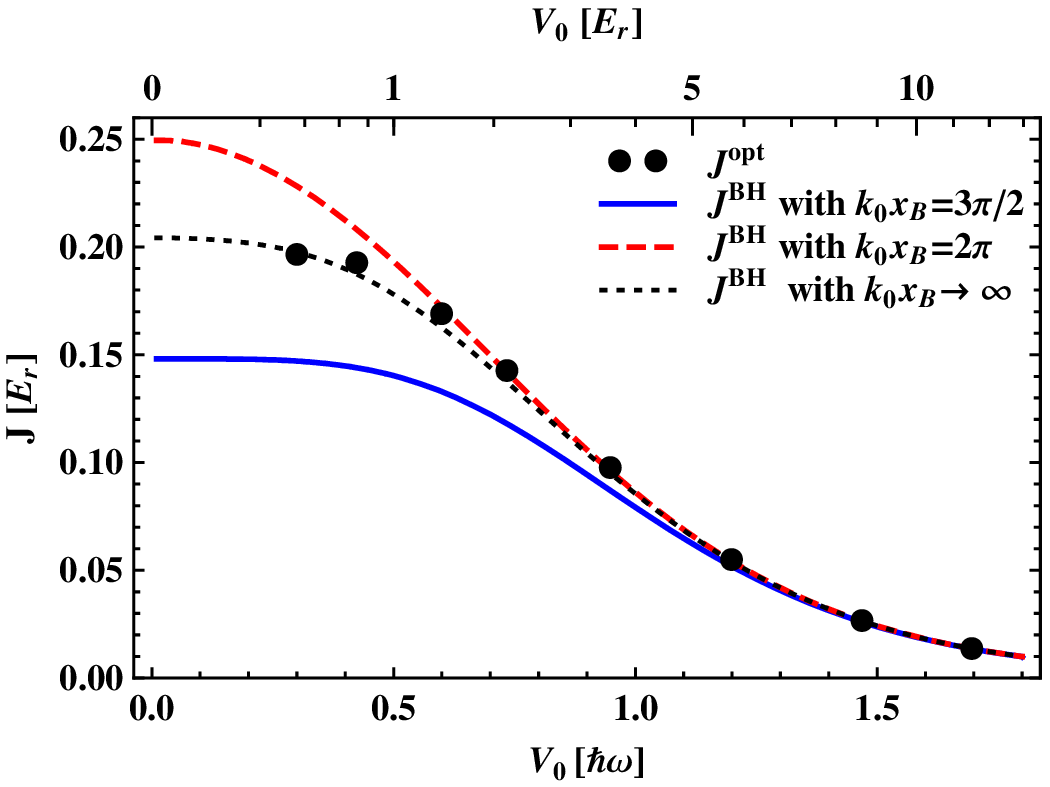}
     \end{minipage}
  {\bf (b)}\begin{minipage}[t]{0.45\textwidth}
     \vspace{0cm} \includegraphics[width=\textwidth]{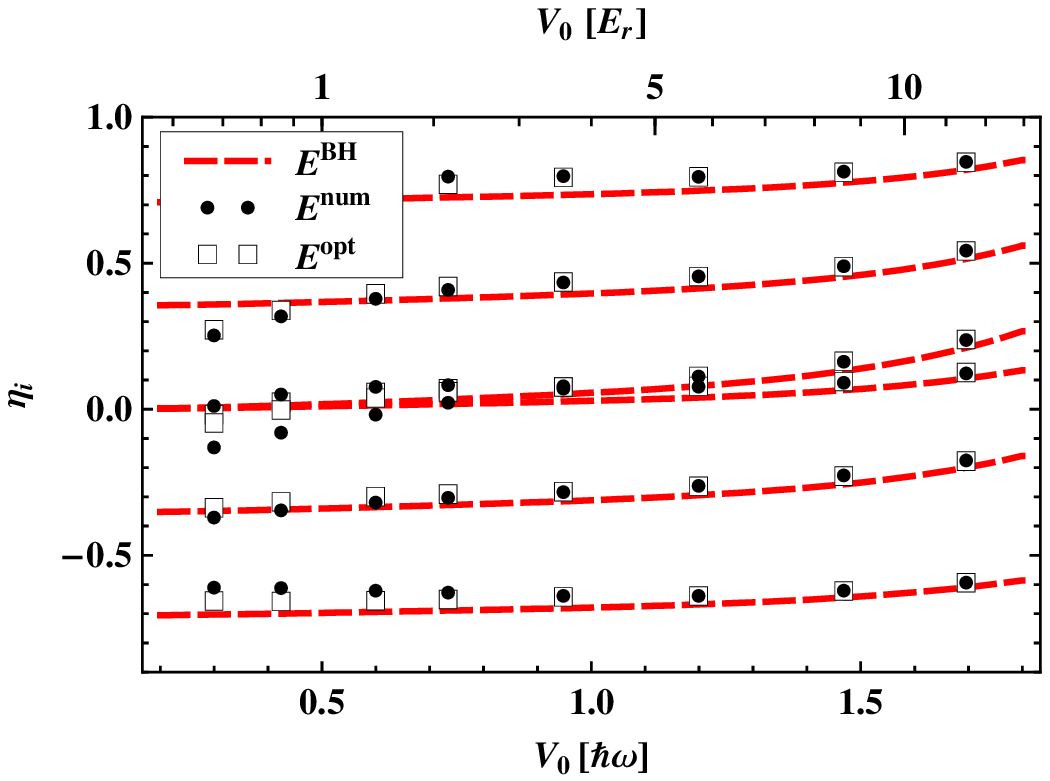}
     \end{minipage}
 \caption{{\footnotesize  (Color online) {\bf (a)} Hopping parameter $J$ as a
  function of the lattice depth $V_0$: the optimal value together with the 
  BH representation for $k_0x_B=\frac{3}{2}\pi, 2\pi, \infty$.
  {\bf (b)} 
  The six eigenenergies in the rescaled form (see
  Eq.~(\ref{eq:scaledEnergyNoInt})) for 
  $a_{\rm sc}=0$ and $U=0$ as a function of the lattice depth $V_0$.
  The BH energies are obtained for onsite energies with boundaries at
  $k_0x_B=3\pi/2$ and a hopping parameter with boundaries at $k_0 x_B = 2\pi$. 
}}
\label{fig:hopping}
\end{figure}

  Since the BH representation of both the hopping and the onsite-energy parameter is
  converged for $V_0>1.5\,\hbar\omega$, the BHM predicts the correct
  eigenenergies in this regime. The mean deviation from the numerical energies 
  for $V_0=1.7\,\hbar\omega$ is $0.25 E_r$ or $0.01\%$ respectively.
  Fig.~\ref{fig:hopping} (b) shows the spectrum of the full numerical calculation.
  In order to compensate for an increase of the 
  energies of the order of $2\epsilon_0$ and the shrinking width 
  of the first Bloch band of the order of $4J$ as $V_0$ is
  increased, rescaled energies $\eta_i$ with
 \begin{equation}
 \label{eq:scaledEnergyNoInt}
   \eta_i(V_0) = \frac{E_i-2\epsilon_0^{\rm BH}}{4 J^{\rm BH}}
 \end{equation}
  are shown. For $V_0>1.5\,\hbar\omega$ the spectrum of the BH
  Hamiltonian in the BH representation is in 
  very good agreement with the spectrum obtained from a full numerical calculation.
  With the optimal parameter set the BHM is able to predict the numerical 
  eigenenergies already for $V_0>\hbar\omega$ with high accuracy 
  (mean relative error in energies $\approx 0.001\%$).
  For shallower lattices numerical and optimal eigenenergies do not match, 
  which means that the BHM is principally unable to reproduce the
  eigenenergies of the triple well for $V_0<\hbar\omega$. 
  Two differences are most obvious: (i) Both the BHM in the BH representation and
  with the optimal BH parameters underestimate significantly the value of the
  ground-state energy while especially the BH representation tends to
  overestimate all other eigenenergies in
  the limit of vanishing lattice depth. (ii) For $U=0$ the BHM  
  predicts an energy difference of $\epsilon_1 - \epsilon_0$ between
  the third and fourth energy level. In the BH representation this difference goes
  to zero for $V_0\rightarrow 0$ (see Fig.~\ref{fig:onsite}) and the levels
  become degenerate. Although this degeneracy is lifted by using the optimal 
  parameter set, the energy difference is still largely underestimated.

  These disagreements can be caused by the dominance of the confinement potential
  over the OL potential for small lattice depths. 
  Also hopping to non-neighboring wells can become important as shown in 
  Sec.~\ref{ssec:EBHM}. 
  As already mentioned, one can assume that the BHM converges better for
  larger lattices with a shallow harmonic external confinement. On the other hand, the
  convergence behavior for, e.\,g., double wells as they appear in superlattices should
  be similar or even worse compared to the case of the triple well.

\subsection{Interaction parameter}
\label{subsec:interaction}

The interaction parameter $U$ depends on both the scattering length $a_{\rm sc}$
and the lattice depth $V_0$ that determines how strong two particles are
confined in one well. First, the dependence on the lattice depth for the case
of a relatively small and a relatively large negative scattering length will be
discussed.

\begin{figure}[t]
 \centering
   {\bf (a)}\begin{minipage}[t]{0.45\textwidth}
     \vspace{0cm}\includegraphics[width=\textwidth]{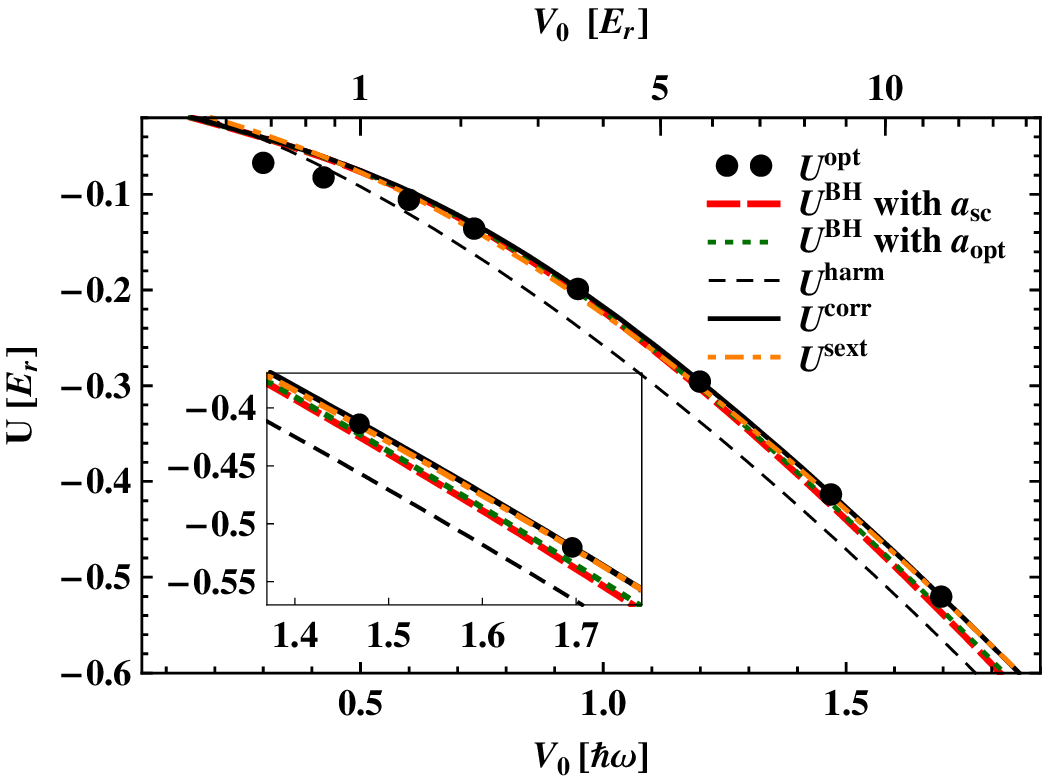}
     \end{minipage}
     {\bf (b)}\begin{minipage}[t]{0.45\textwidth}
     \vspace{0cm}\includegraphics[width=\textwidth]{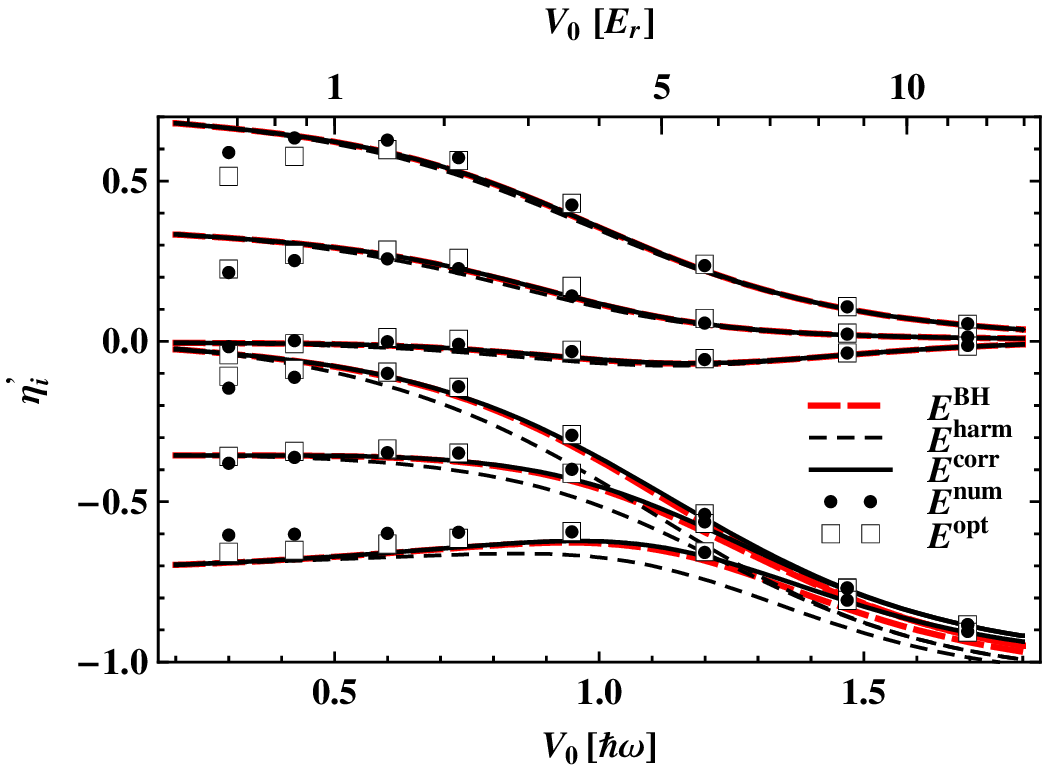}
     \end{minipage}
\caption{{\footnotesize (Color online) {\bf (a)} The Interaction parameter as
  a function of the lattice depth $V_0$ in the weak interacting regime with 
  $a_{\rm sc}\approx -180 a_0 $. 
  This results in $0.02\leq\left|a_{\rm sc}/a_{\rm ho}\right|\leq 0.08$ in this graph. 
 {\bf (b)} The six eigenenergies in the rescaled form (see
  Eq.~(\ref{eq:scaledEnergyInt})) as a function of the lattice depth $V_0$.
}}
\label{fig:interaction_weak}
\end{figure}

Fig.~\ref{fig:interaction_weak} (a) shows a plot of the optimal interaction
parameter $U^{\rm opt}$ together with several predictions of the interaction energy by
$U^{\rm BH},U^{\rm harm}$ and $U^{\rm corr}$. The scattering length is fixed to
about $- 180 a_0 $. For this value, the ratio $\left|a_{\rm sc}/a_{\rm ho}\right|$ 
which increases with the lattice depth $V_0$ reaches up to only $0.08$ in this graph. 

Within the BHM the interaction parameter $U^{\rm BH}$ is independent of the 
confining potential $\hat V_{\rm conf}$.
Accordingly, the main source of error is not that the Wannier
functions are not adapted to the confining potential, but that they do not
reflect effects of interaction.
Thus, for weak interaction the BH representation predicts $U^{\rm opt}$
quite well already for shallow lattices with $V_0\approx\hbar\omega$. 
However, an increasing value of $V_0$  amplifies the confinement and thus the ratio
$\left|a_{\rm sc}/a_{\rm ho}\right|$. Remarkably, this causes a small error of 
$U^{\rm BH}$ which is already visible for $\left|a_{\rm sc}/a_{\rm ho}\right|<0.08$ 
and reduces the validity regime of $U^{\rm BH}$ to scattering lengths which are 
at least one order of magnitude smaller than the trap length $a_{\rm ho}$.

For weak interaction the energy-offset in the harmonic well $U^{\rm harm}$ can
predict $U^{\rm opt}$ only poorly compared to $U^{\rm BH}$.
On the other hand, the corrected analytical approach $U^{\rm corr}$ introduced in 
Sec.~\ref{ssec:Ucorr} is in 
excellent agreement with the optimal interaction parameter for $V_0>\hbar\omega$.
This is also reflected by the energy spectrum shown in
Fig.~\ref{fig:interaction_weak} (b).

In Fig.~\ref{fig:interaction_weak} (b) the upper three states describe particles 
in different wells with energies almost independent of the interaction.
The first three states, on the other hand, describe attractively interacting particles 
in the same well. Within the BHM their energies are shifted in the order of $U$. 
Thus, the rescaling of the energies (Eq.~\ref{eq:scaledEnergyNoInt}) is
changed to
 \begin{equation}
 \label{eq:scaledEnergyInt}
   \eta_i'(V_0) = \frac{E_i-2\epsilon_0^{\rm BH}}{4 J^{\rm BH}+|U^{\rm corr}|}.
 \end{equation}

\begin{figure}[t]
 \centering
   {\bf (a)}\begin{minipage}[t]{0.45\textwidth}
     \vspace{0cm}\includegraphics[width=\textwidth]{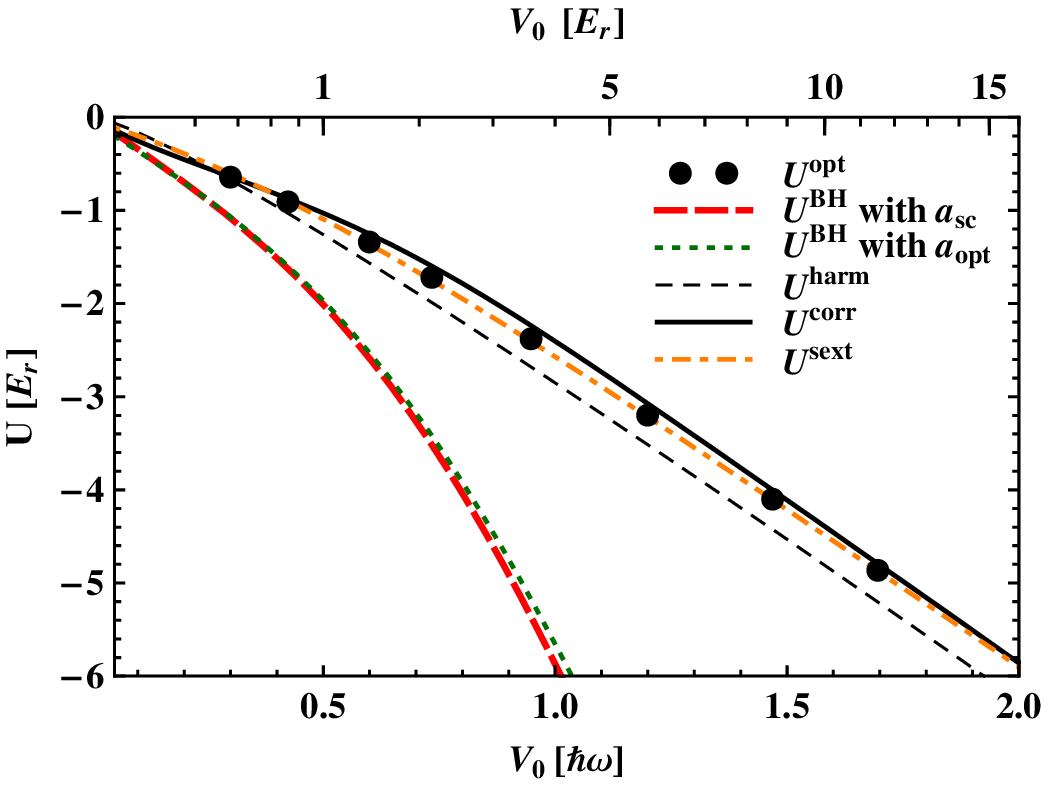}
     \end{minipage}
  {\bf (b)}\begin{minipage}[t]{0.45\textwidth}
     \vspace{0cm}\includegraphics[width=\textwidth]{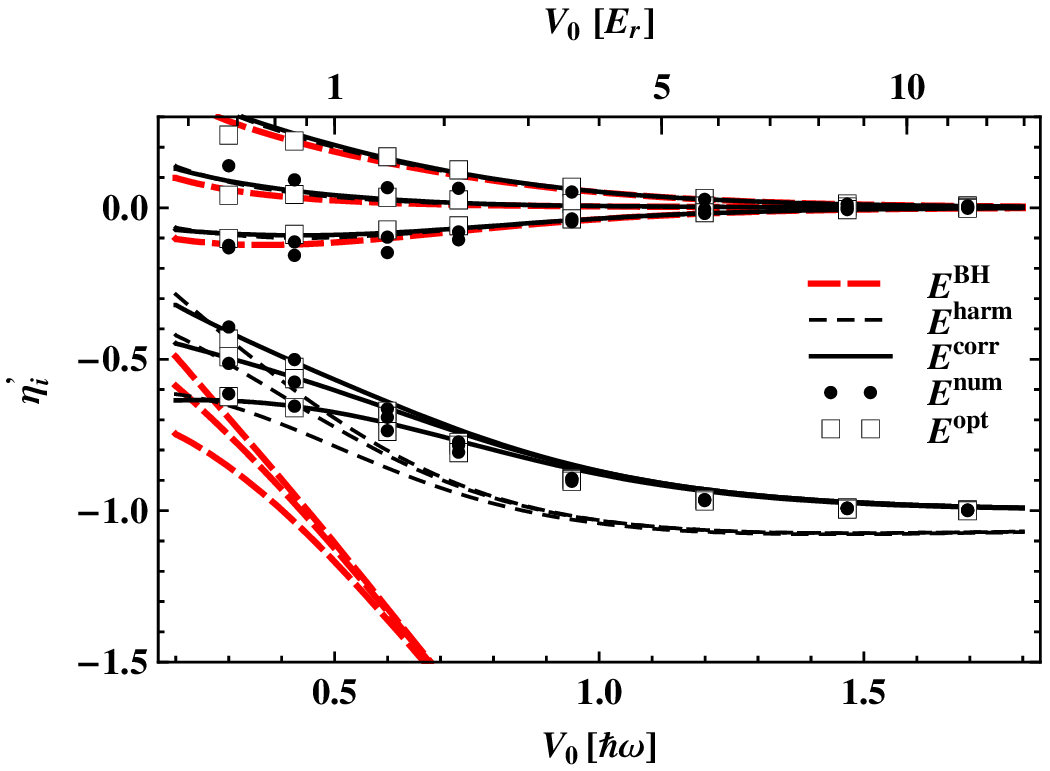}
     \end{minipage}
\caption{{\footnotesize (Color online) {\bf (a)} The Interaction parameter as
  a function of the lattice depth $V_0$ in the strong interacting
  regime with $a_{\rm sc}\approx -4600 a_0$. 
  This results in $0.5\leq\left|a_{\rm sc}/a_{\rm ho} \right|\leq 2.2$ 
  in this graph. 
  {\bf (b)} The six eigenenergies in the rescaled form (see
  Eq.~(\ref{eq:scaledEnergyInt})) as a function of the lattice depth $V_0$.
}}
\label{fig:interaction_strong}
\end{figure}

The energies of the lowest three states are well
predicted by using $U^{\rm corr}$ for $V_0\geq 1.2\,\hbar\omega$ while
both the BH representation and the harmonic approximation underestimate the
energies more and more as the lattice depth and with this $|a_{\rm sc}/a_{\rm
  ho}|$ increases. For lattice depths smaller than $\hbar\omega$
the errors from the non-interacting case are inherited, e.\,g. the ground-state 
energy and the splitting between the third and the fourth
level is underestimated by all models.

In Fig.~\ref{fig:interaction_strong} (a) the interaction parameter is plotted 
for a large negative scattering length of about $-4600 a_0 $ which means, that
the absolute value of the scattering length is up to two times larger than the 
trap length $a_{\rm ho}$.
As the magnitude of $a_{\rm sc}$ is of the order of $a_{\rm ho}$, the BH 
representation of $U$ differs drastically from the real interaction parameter.
In this regime the linear dependence of $U^{\rm BH}$ on the
scattering length leads to significantly wrong results (see the inset in 
Fig.~\ref{fig:Corr}). 
In this case, the energy offset $U^{\rm harm}$ provides a far better
approximation, but it results in an almost constant underestimation of 
the optimal interaction strength as is evident from 
Fig.~\ref{fig:interaction_strong} (a). 
The corrected interaction parameter, on the other hand, gets closer to $U^{\rm
  opt}$ for deep lattices. The graph supports that
Eq.~(\ref{eq:validity_of_correction}) indeed reflects the validity regime of $\A$.
As $U$ is approximately proportional to $V_0/\hbar\omega$ the ratio
$V_0/|U^{\rm  corr}|$ shrinks with an increasing lattice depth like
$(\hbar\omega)^{-1}$ and $U^{\rm corr}$ approaches the optimal interaction parameter.
The same is again reflected by the energy spectrum shown in
Fig.~\ref{fig:interaction_strong} (b). Only $U^{\rm corr}$ is able to
reproduce the numerical eigenenergies of the three lowest states in deep
lattices, while $U^{\rm harm}$
generates an almost constant underestimation and $U^{\rm BH}$ a rapidly
increasing underestimation of the numerical eigenenergies of these states. 

\subsubsection{Sextic approximation}

Figs.~\ref{fig:interaction_weak} (a) and \ref{fig:interaction_strong} (a)
show additionally the estimate of the interaction parameter using $U^{\rm sext}$, the
interaction energy obtained in the sextic trap. 
It agrees precisely with the optimal parameter $U^{\rm opt}$ independently of the
strength of interaction provided that $V_0>1.5\,\hbar\omega$. Also for shallower
lattices, it is close to the optimal value. 
Thus, the examination of the interaction in a sextic trap 
\cite{cold:gris09} can already cover all anharmonic features of the 
onsite interaction for sufficiently deep lattices. 
This implies that one can treat the anharmonic form of a lattice site as
a perturbation to the harmonic approximation to study interactions in OLs
more realistically. For example, in \cite{cold:ment09} this has been done by 
treating the difference between the
OL potential and the harmonic approximation as a perturbing potential.
The effects of the anharmonicity within a single site of an optical lattice
have also been investigated non-perturbatively in \cite{cold:deur08}.

\subsubsection{Energy-dependent scattering length}

One could expect that the use of the optimal energy-dependent scattering length
$a_{\rm opt}$ instead of $a_{\rm sc}$ can improve the value of $U^{\rm BH}$
(see Eq.~(\ref{eq:U_BH})) significantly.
This is, however, not the case, as can be clearly seen in
Fig.~\ref{fig:interaction_weak} (a) and Fig.~\ref{fig:interaction_strong} (a).
In the validity regime of the BH representation ($|a_{\rm sc}|\ll a_{\rm ho}$)
the energy dependence of the scattering length seems not to be of importance. 
Also for strong interaction the offset between $U^{\rm BH}$ when used with $a_{\rm
  sc}$ or $a_{\rm opt}$ is small compared to the offset between any of the
$U^{\rm BH}$ and $U^{\rm opt}$.
In other words, the effect of the energy dependence of the
scattering length in the OL is generally small compared to the error due to
the use of the interaction-independent Wannier basis.

To quantify the effect of the energy-dependence, we use 
\begin{equation}
 \Delta U=
  \frac{
      U^{\rm harm}(a_{\rm sc})-U^{\rm harm}(a_{\rm opt})
  }{
      U^{\rm harm}(a_{\rm opt})
  },
\end{equation} 
the error of the interaction energy in the harmonic well, 
if $a_{\rm sc}$ is used instead of $a_{\rm opt}$. This estimates the error of 
neglecting the energy dependence of the scattering length independently of anharmonic
effects.
For a sufficiently large value of $|a_{\rm sc}/a_{\rm ho}|$ the deviation of
$U^{\rm harm}$ from the linear behavior of $U^{\rm BH}$ (see inset of
Fig.~\ref{fig:Corr}) starts to be significant. Therefore, the size of 
$a_{\rm sc}$ relative to $a_{\rm opt}$ does not directly show the impact of the
energy dependence.
 
In Tab.~\ref{tab:errors} besides $\Delta U$
the errors of $U^{\rm BH}$, $U^{\rm harm}$ and $U^{\rm corr}$ compared to the
optimal value are listed for various scattering lengths and lattice depths. 
For very deep lattices with $V_0>12 E_r$ where a full numerical
calculation in the triple well is too laborious we use the sextic
approximation $U^{\rm sext}$ instead of the optimal interaction parameter,
as it exactly reproduces $U^{\rm opt}$ for $V_0>10 E_r$.

The value of $\Delta U$ in Tab.~\ref{tab:errors} shows to
depend mainly on the ratio $a_{\rm sc}/a_{\rm ho}$. However, even for a large ratio it
amounts up to only a few percent. Also the error of $U^{\rm BH}$ depends on that
ratio, but grows much faster with $|a_{\rm sc}/a_{\rm ho}|$. 
This reflects again the complete breakdown of the BH representation for 
scattering lengths with a magnitude of the order of the trap length.

The error of $U^{\rm harm}$ decreases, though quite slowly, with growing
lattice depth $V_0$. It comprises the error of an energy-independent
scattering length ($\Delta U$) 
and of neglecting anharmonic effects. The latter effect can be estimated by subtracting 
the value of $\Delta U$ from the error of 
$U^{\rm harm}$. 
Only for the largest depth considered in Tab.~\ref{tab:errors} ($V_0=64.7
E_r$) effects of $\Delta U$ are bigger than anharmonic effects.
For a fixed lattice depth $V_0=11.5E_r$ and a decrease of 
the scattering length from $-0.08 a_{\rm ho}$ to $-2.01 a_{\rm ho}$ anharmonic 
effects seem to become less important. This can be explained by the stronger confinement 
in the wells due to the attractive interaction.

The error of $U^{\rm corr}$ results from the error of the correction factor
$\A$ and from the error $\Delta U$. The error of $\A$ can again be estimated
by subtracting $\Delta U$ from the error of 
$U^{\rm corr}$. This shows that the error of $\A$ depends, as expected, 
on the ratio $U^{\rm corr}/V_0$. 
In all considered cases it is found to be smaller than 
the error of $U^{\rm harm}$ caused by the anharmonicity and possesses 
the opposite sign of $\Delta U$. 
Therefore, the total error of $U^{\rm corr}$ is generally
much smaller than the error of $U^{\rm BH}$ or $U^{\rm harm}$.

\begin{table}[ht]
\caption{The error $\Delta U$ of the use of an energy independent
scattering length together with errors of $U^{\rm BH}$, $U^{\rm harm}$ and $U^{\rm corr}$ relative 
to $U^{\rm opt}$ for various lattice depths and scattering lengths.}

 \centering
\begin{tabular}{l c c c c}
\hline\hline
$V_0/E_r$ &                             11.5    & 11.5   & 25.2   & 64.7\\
$a_{\rm sc}/a_{\rm ho}$ &                -0.08   & -2.01  & -2.44   & -3.09\\
$U^{\rm corr}/V_0$ &                     -0.05   & -0.42  &  -0.30 & -0.20\\[0.5ex]
\hline
$\Delta U$ & -0.58\% & -1.5\% & -1.8\% & -2.2\%\\ 
\hline
Error of $U^{\rm BH}$ &                 -3.18\% & -190\%  & -243\% & -320\%\\
Error of $U^{\rm harm}$ &               -8.53\% & -7.0\%   & -5.1\%   &  -4.2\%\\
Error of $U^{\rm corr}$ &                0.06\% & 1.5\%  & 0.04\% & -1.1\%\\ 
\hline\hline
 \end{tabular}
 \label{tab:errors}
\end{table}

\subsubsection{Interaction energy at a resonance of the scattering length}

\begin{figure}[t]
 \centering
   {\bf (a)}\begin{minipage}[t]{0.45\textwidth}
     \vspace{0cm}\includegraphics[width=\textwidth]{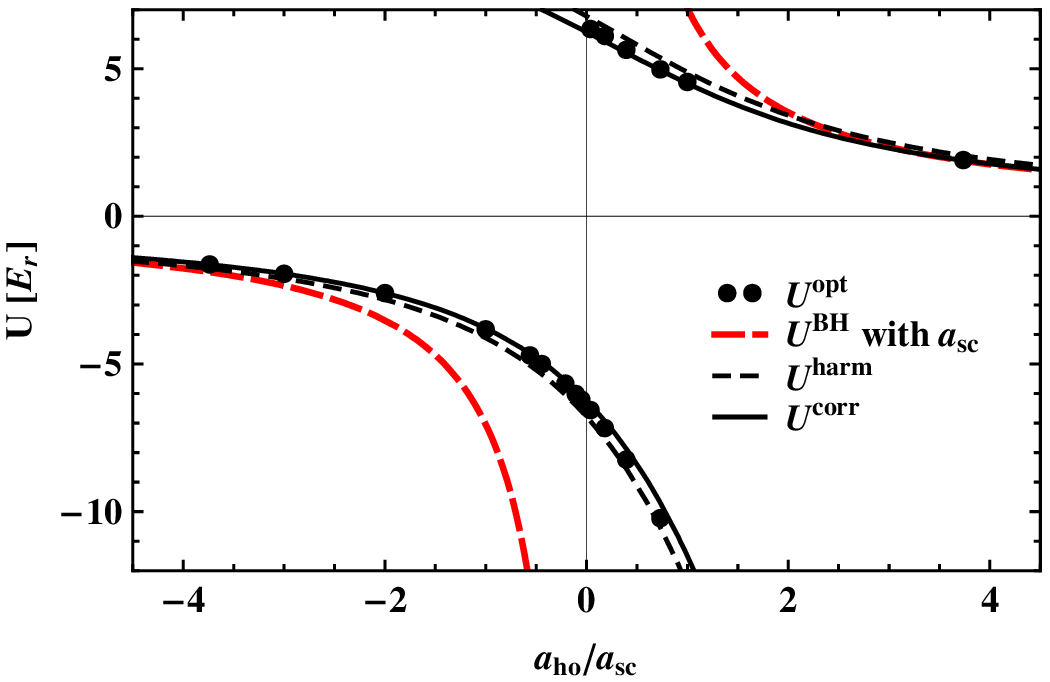}
     \end{minipage}
  {\bf (b)}\begin{minipage}[t]{0.45\textwidth}
     \vspace{0cm}\includegraphics[width=\textwidth]{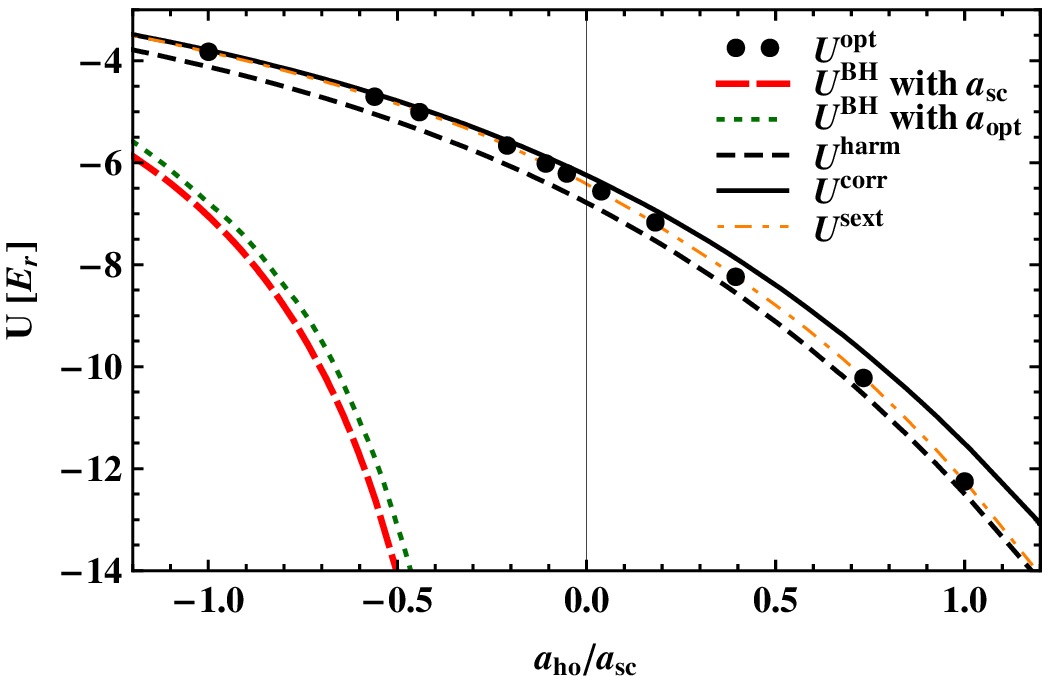}
     \end{minipage}
\caption{{\footnotesize (Color online) The Interaction parameter for different scattering
    lengths at $\frac{V_0}{\hbar \omega}=1.7$. To observe the behavior at the
  resonance, the dependence on $\frac{a_{\rm ho}}{a_{\rm sc}}$ is shown.}}
\label{fig:U_vs_Asc}
\end{figure}

Until now the two cases of very small and very large interaction were
discussed.
To examine in more detail the regime of validity for $U^{\rm BH}$, $U^{\rm
  harm}$ and $U^{\rm corr}$
we consider in Fig.~\ref{fig:U_vs_Asc} the behavior of the interaction
parameter at a resonance ($a_{\rm ho}/a_{\rm sc}\to 0$) for a lattice depth of
$V_0=1.7\,\hbar \omega$. At this depth the BH representation
of the other parameters ($J,\epsilon_0$ and $\epsilon_{\pm1}$) is close to 
their optimal values.

Since at a resonance the energy regime of higher Bloch bands can be reached the
notion of Bloch bands should be reconsidered in the presence of the new interaction
parameters $U^{\rm harm}$ and $U^{\rm corr}$. 
For strong interaction one normally speaks of a coupling to higher Bloch bands,
which means, that the true eigenfunctions can only be reproduced by an expansion
into Wannier functions of several Bloch bands.
The new interaction parameters are, however, not formulated as matrix elements
in the Wannier basis. The resulting BH Hamiltonian is therefore intrinsicely a
multi-band Hamiltonian. As shown in the last paragraphs the use of an optimal
parameter set can reproduce the numerical energy spectrum very well. One can 
therefore describe multi-band physics surprisingly well within the simple BHM.
As we will show now this has, however, limitations for repelling
particles where the BH states become degenerate with excited
``molecular'' states and noninteracting ones in higher Bloch bands.

First of all, by varying the scattering length over the resonance ($a_{\rm
  ho}/a_{\rm sc}=0$) it becomes obvious, that the BH representation $U^{\rm BH}$
is not valid in the proximity of the resonance where it diverges to
$\pm \infty$ (see Fig.~\ref{fig:U_vs_Asc} (a)).
On the other hand, $U^{\rm corr}$
approximates well the optimal value $U^{\rm opt}$ for a large range of 
repelling (upper branch in Fig.~\ref{fig:U_vs_Asc} (a)) and
attracting (lower branch in Fig.~\ref{fig:U_vs_Asc} (a)) interactions. 

In Fig.~\ref{fig:U_vs_Asc} (b) the lower branch is shown on an enlarged scale 
around the resonance. It shows that $U^{\rm corr}$ is valid until 
$|a_{\rm sc}|\approx a_{\rm ho}$.
For growing $|U|$ the correction factor $\A$ looses slowly its
validity. Again, one can observe that the behavior of $U^{\rm opt}$ is
exactly followed by the sextic approximation while the harmonic
approximation generally overestimates the magnitude of $U^{\rm opt}$.
  
\begin{figure}[t]
 \centering
  {\bf (a)}\begin{minipage}[t]{0.45\textwidth}
     \vspace{0cm}\includegraphics[width=\textwidth]{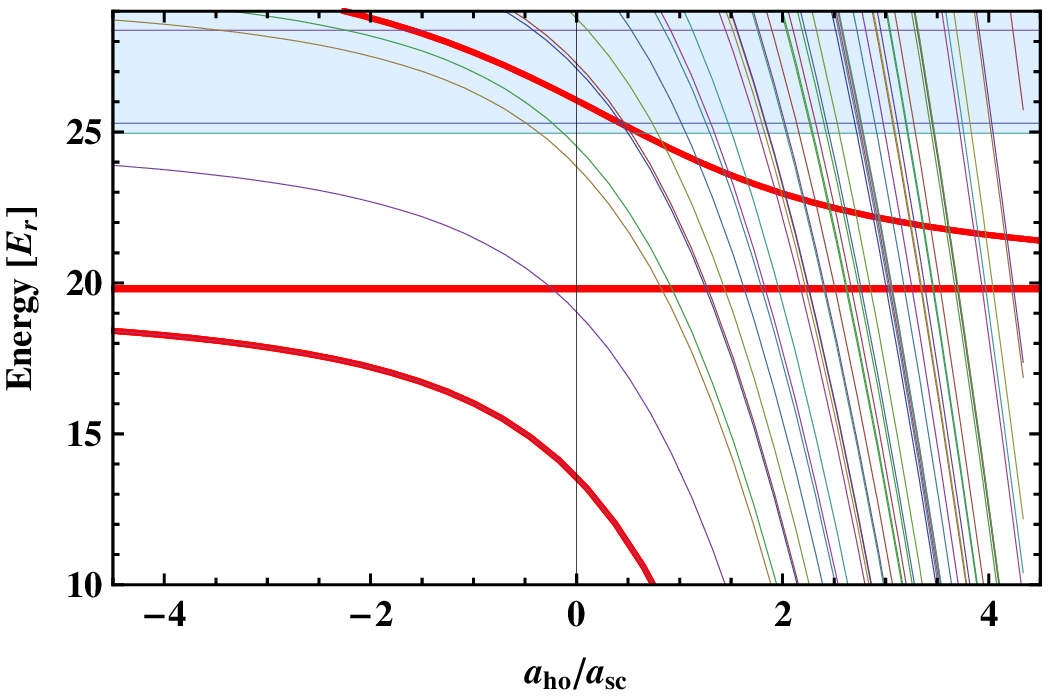}
   \end{minipage}
  {\bf (b)}\begin{minipage}[t]{0.45\textwidth}
     \vspace{0cm}\includegraphics[width=\textwidth]{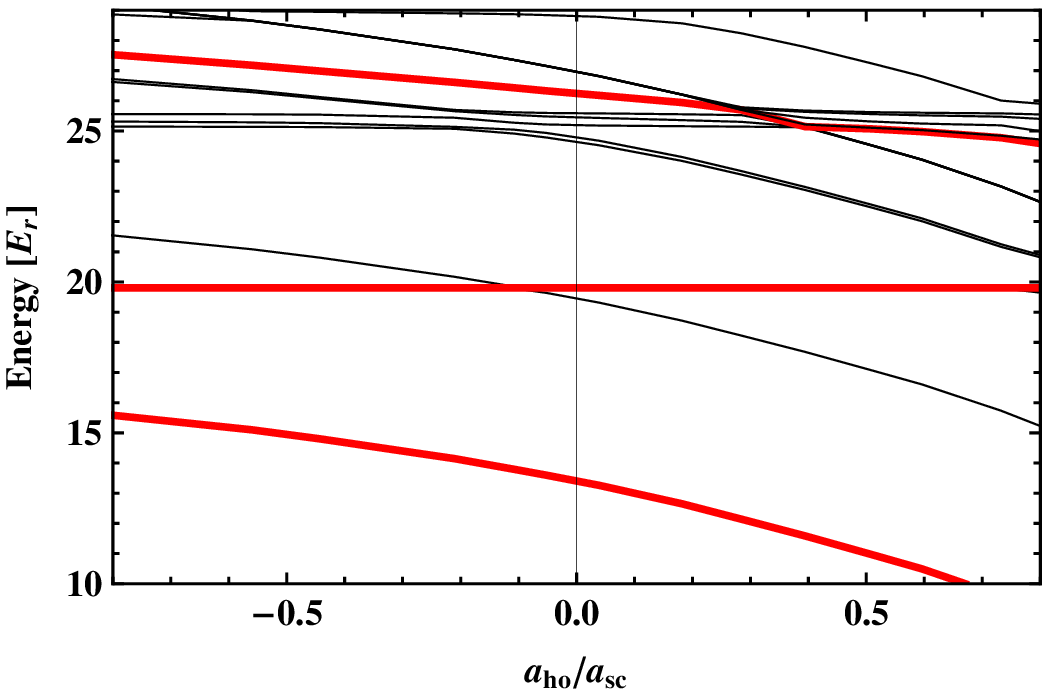}
     \end{minipage}
 \caption{{\footnotesize (Color online) Eigenenergies as a function of the
     scattering length at $\frac{V_0}{\hbar \omega}=1.7$. 
     {\bf (a)} Sketch of energy spectrum (see text for details).
     {\bf (b)} Energy spectrum obtained from a full numerical calculation. The
     three thick (red) lines indicate energies which are closest to the energies predicted
     by the BHM.}}
 \label{fig:energy_crossing}
\end{figure}

On the upper branch in Fig.~\ref{fig:U_vs_Asc} near the resonance 
the energy of bound particles in higher Bloch bands can cross the BH states.
To understand how this happens, we have sketched in
Fig.~\ref{fig:energy_crossing} (a) a realistic estimate of a part of the energy
spectrum.
The spectrum has mainly three constituents: {\bf (i)} The thick (red) lines represent the
eigenvalues of the BHM of the first Bloch band (with $U=U^{\rm corr}$). 
The lower branch consists
of three states with attracting particles ($U<0$) in the same well, and the
upper branch of three states with repelling particles  ($U>0$) in the same
well. In the horizontal middle
branch are all remaining energies with particles in different wells. These
states are only weakly influenced by the interaction. 
{\bf (ii)} The bunch of thin lines going from top to bottom represent
``dimers'' similar to the lowest BH states but in higher Bloch bands.
Their energy consists of single-particle energies of Bloch states in different
bands and the interaction energy $U^{\rm corr}<0$.
{\bf (iii)} The thin horizontal lines at the top represent the energies of
non-interacting states where one particle is in the first Bloch band and one
in the second Bloch band of at least one direction. The light (blue) filling 
indicates that the energy regime of higher Bloch bands is reached. 
Note, for any depth of the lattice two repulsively interacting
particles will enter this energy regime at the resonance as it is always 
half way to the energy of two particles in the second Bloch band.

In Fig.~\ref{fig:energy_crossing} (b) we have plotted in the region of resonance
the CI eigenenergies of the lowest totally symmetric states in the trap. 
As they are of the same symmetry, avoided crossings occur. These appear close to 
where the estimated energy spectrum (Fig.~\ref{fig:energy_crossing} (a))
shows (true) energy crossings.  
The avoided crossings are caused by a more or less strong coupling between states in
different Bloch bands, which is not incorporated in the estimated spectrum.
Particularly, the energy of states which can be identified with BH states of
repulsively interacting particles is distorted due to avoided crossings. 
Thus, one cannot speak of pure repulsively interacting states in the first
Bloch band any more. In contrast, the coupling between non-interacting BH
states and the first exited ``dimer'' is very small and no energy distortion
is visible at the resolution of Fig.~\ref{fig:energy_crossing} (b).

The complicated level structure indicates that the BHM, even if  
$U^{\rm corr}$ is used, has to be handled 
with care, if repulsively interacting and also non-interacting states at a
resonance are to be described. In its corrected form it is, however, 
useful as a first approximation to find level crossings.

The rich structure of avoided crossings offers potentially a scheme to convert 
states of repulsively interacting, attractively interacting, and
non-interacting nature into each other by performing sweeps of the scattering length. 
To estimate the dynamical behavior of the system the knowledge of the 
coupling strengths between states in different Bloch bands is crucial.
Since these couplings are ignored within the BHM more sophisticated calculations 
such as the one presented in this work have to be performed to gain a deeper
insight.
 
\subsection{Extended Bose-Hubbard model}
\label{ssec:EBHM}
Especially for shallow lattices
both the interaction between particles in nearest-neighbor (NN) wells and hopping to 
next-to-nearest neighbors (NtNN) can become important 
\cite{cold:mazz06,cold:boer07,cold:trot08}. 
One can easily account for this by including the according matrix elements from 
$\hat H_{\rm Wan}$ of Eq.~(\ref{eq:HamWan}) into an
extended Bose-Hubbard model (EBHM) with Hamiltonian
\begin{equation}
\begin{split}
\hat H_{\rm EBH} 
  =&\, \hat H_{\rm BH} 
    + J_2 (\hat b_{-1}^\dagger \hat b_{1} +  \hat b_{1}^\dagger \hat b_{-1}) \\ 
  & + U_1 \sum_{<i,k>}\left( 
   \hat b^\dagger_i \hat b_i^\dagger \hat b_i \hat b_k +
   \hat b^\dagger_i \hat b_k^\dagger \hat b_k \hat b_k \right)
   \\
  &+ U_2 \sum_{<i,k>}\left( 
   \hat b^\dagger_i \hat b_i^\dagger \hat b_k \hat b_k +
   \hat b^\dagger_i \hat b_k^\dagger \hat b_i \hat b_k \right)\, . \end{split}
\label{eq:EBHHam}
\end{equation} 
The NtNN hopping parameter
$J_2 = \left<w_{-1}\right|\frac{\hat p^2}{2m} + \hat V_{\rm OL}\left|w_{1}\right>$
decays much faster with the lattice depth than $J$. For $V_0=\hbar\omega$ it is one 
order smaller than $J$ and for $V_0=1.5\,\hbar\omega$ already two orders.
Furthermore, the EBHM possesses in BH representation two new parameters 
$U_1 = \frac{4\pi \hbar^2 a_{\rm sc} }{m}\int d^3\vec r\, w_0(\vec r\,)^3 w_1(\vec r\,)$,
and 
$U_2 = \frac{4\pi \hbar^2 a_{\rm sc} }{m}\int d^3\vec r\, w_0(\vec r\,)^2 w_1(\vec r\,)^2$,
describing NN interaction. These parameters are generally at least one order of magnitude 
smaller than $U$.
Unfortunately, the BH representation of $U_1$ and $U_2$ cannot be improved like
for $U$ as there exists no analytical solution for two interacting particles in
different wells. 
Therefore, only if the following conditions are met, the EBHM can be quantitatively 
better than the BHM: 
{\bf (i)} The scattering length is small compared
to the trap length (i.\,e. $|a_{\rm sc}|\ll a_{\rm ho}$), since otherwise $U_1$ and $U_2$
are predicted inaccurately.
{\bf (ii)} The lattice is deep enough that
errors of the BH parameters are smaller than $U_1$, $U_2$ or $J_2$.

Fulfilling the first condition, the case of weak interaction ($a_{\rm sc}=-176\,a_0$)
is discussed in the following.
The second condition is not easy to meet in few-well systems, since especially 
the BH representation of the onsite-energies reflects the optimal parameters
only for deep lattices (see the inset in Fig.~\ref{fig:onsite}), 
where $U_1$, $U_2$ and $J_2$ are already two orders of magnitude smaller 
than $U$ and $J$, respectively.

However, many observables do not depend on the absolute value of the eigenenergies, 
but on their differences. Energy differences determine for example the characteristic 
frequencies of a state in a coherent superposition of different eigenstates.
These frequencies can be measured experimentally with high accuracy such that deviations 
of the simple BHM can be observed \cite{cold:trot08}.

The influence of the errors of the onsite energies can be reduced by considering 
differences of eigenenergies as the latter do not depend on the absolute value of the 
on-site energies but only on the offset $\Delta=\epsilon_1-\epsilon_0$.
Although $\Delta$ is also predicted inaccurately in the BH representation 
for $V_0 < 1.2\,\hbar\omega$, it has often only a minor influence on the energy differences.
This is especially the case for the energy difference $E_5-E_2$ for which 
a diagonalization of the BH Hamiltonian without interaction yields 
\begin{equation}
E_5-E_2 = \sqrt{J^2+\Delta^2}\,.
\end{equation} 
For the present potential with $V_0 < 1.2\,\hbar\omega$ the relation $J^2 \gg \Delta^2$ 
holds. Since the errors of the parameters $J$ and $U$ are small for lattice depths down 
to $0.7\,\hbar\omega$, one can clearly identify effects of NN interaction and NtNN hopping 
as proper corrections of the BHM for this energy difference.


%
%

\begin{figure}[ht]
 \centering
   \begin{minipage}[t]{0.45\textwidth}
     \vspace{0cm}\includegraphics[width=\textwidth]{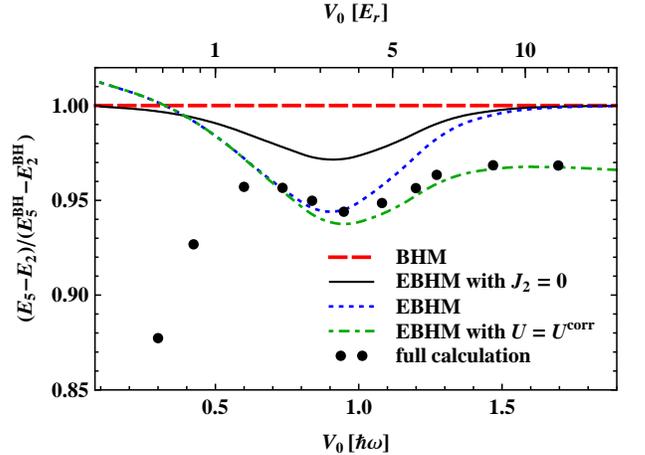}
     \end{minipage}
\caption{{\footnotesize (Color online) Energy difference $E_5-E_2$ in different orders 
  of approximation as a function of the lattice depth $V_0$ for 
  $a_{\rm sc}\approx -180 a_0$. 
  The value of $E_5-E_2$ is shown relative to the same energy 
  difference predicted by the BHM in BH representation.
  The BH representation of $\epsilon_0,\epsilon_{\rm \pm1}$ was determined for 
  $k_0 x_B = 3\pi/2$ and the value of $J$ and $J_2$ for $k_0 x_B = 2\pi$.}}
\label{fig:BH_vs_EBH}
\end{figure}

In Fig.~\ref{fig:BH_vs_EBH} $E_5-E_2$ is shown relative to its value for
the BHM with parameters steming from the BH representation.
The correct energy difference is compared to the one predicted by the BHM 
which is successively amended.
First, only effects of NN interaction are incorporated by the use of the EBHM without 
NtNN hopping (i.\,e. setting $J_2=0$). Then the full EBHM is considered which is finally 
corrected by replacing $U^{\rm BH}$ by $U^{\rm corr}$.
The relative importance of these three corrections for different lattice depths becomes 
clearly visible. Due to the weakness of the interaction NtNN hopping and NN interaction have 
comparable influence, most dominant at $V_0\approx 0.9\,\hbar\omega$.
For deeper lattices the correction of the interaction parameter becomes increasingly
important and dominates finally for $V_0 > 1.3\,\hbar \omega$ the correction of the BHM.

By including all three corrections one is able to determine the correct energy difference 
with about 1\% accuracy down to lattice depths of only $V_0 = 0.7\,\hbar\omega$.
Below this lattice depth the EBHM is unable to improve the BHM significantly.
Additional effects steming from the confinement $\hat V_{\rm conf}$ finally dominate
the properties of the energy spectrum and the description by a BHM or even an EBHM 
becomes inappropriate.

Also for other differences of eigenenergies the prediction of the EBHM improves the
one of the BHM at least qualitatively (depending on the importance of $\Delta$).
Only for differences involving the first energy level this is not the case.
Since for attractive interaction (as shown in Fig.~\ref{fig:BH_vs_EBH}) particles in the 
ground state predominantely occupy the middle well, the energy difference to higher 
lying states, where also the outher wells are occupied, is particularly influenced 
by $\Delta$. The error in the determination of $\Delta$ can therefore dominate over 
improvements of the EBHM.
We have verified that the failure of the EBHM to improve the BHM 
results for energy differences involving the ground-state energy 
$E_1$ does not occur, if repulsive interactions are considered.

\section{Conclusions}
%
\label{sec:conc_outlook}
We have investigated the validity of the standard and an extended Bose-Hubbard
model in a triple-well optical lattice.  
We defined the optimal Bose-Hubbard parameter set to be the one which
reproduces best the eigenenergies of our full numerical calculations.
This allowed to determine when the Bose-Hubbard model is principally able to
describe the system under consideration and whether corrections of the standard 
Bose-Hubbard parameters are needed. 

For shallow optical lattices the confinement to three wells had a notable impact 
on the hopping parameter and the onsite energies. This impact could not be fully 
reflected  by the standard values of these parameters. Disagreements due to 
the confinement vanished, however, for lattice depths $V_0$ larger than the 
ground-state energy of the harmonic approximation $1.5\,\hbar\omega$.
Therefore, in this regime our results are valid for both large optical lattices and 
few-well systems. For results concerning the interaction parameter this is also 
true for shallower lattices, since this parameter is barely influenced by the 
confinement.

The standard interaction parameter showed to be valid only when the scattering 
length is smaller than one percent of the trap length.
However, for this parameter we found an analytic expression which approximates 
the optimal interaction parameter very well for a large range of lattice
depths and interaction strengths.
Since its definition 
is quite general, one can expect
that it is extendable to hetero-nuclear systems, interacting particles in 
different Bloch bands, and lattices with different wavelengths in the three
spacial directions.

The error due to the neglect of the energy-dependence of the scattering length 
was quantified and found to be of less importance than effects caused by the 
incompleteness of the Wannier basis. For deep lattices this error is comparable 
to anharmonic effects.

Furthermore, we found that the sextic expansion of one well can already cover the
full onsite interaction in the lattice. 
We have presented and explained the avoided crossing of repulsively
bound pairs with states in higher Bloch bands.
Finally, an extended Bose-Hubbard model was investigated and it was shown that in a
region of small lattice depths it can partially improve the standard Bose-Hubbard 
model significantly.

\section*{Acknowledgments}
%

 The authors are grateful to the {\it Stifterverband f\"ur die Deutsche
 Wissenschaft}, the {\it Fonds der Chemischen Industrie} and the 
 {\it Deutsche Forschungsgemeinschaft} (within {\it Sonderforschungsbereich 
 SFB\,450}) for financial support.


\end{document}